\documentclass[pmlr]{jmlr}

\usepackage{longtable}

\usepackage{booktabs}
\usepackage{siunitx} 
\usepackage{float}

\usepackage{xcolor}

\theorembodyfont{\upshape}
\theoremheaderfont{\scshape}
\theorempostheader{:}
\theoremsep{\newline}

\title[Towards an Efficient, Customizable, and Accessible AI Tutor]{Towards an Efficient, Customizable, and Accessible AI Tutor}

\author{
        \Name{Juan Segundo Hevia\nametag{\thanks{Corresponding Author}}} \Email{jh216@rice.edu}\\
        \addr Computer Science, Rice University, USA \\
        \AND
\Name{Facundo Arredondo} \Email{fa32@rice.edu}\\
        \addr Computer Science, Rice University, USA \\
        \AND
        \Name{Vishesh Kumar} \Email{vk38@rice.edu}\\
        \addr Applied Physics, Rice University, USA
}

\begin{document}

\maketitle

\vspace{-3em}
\begin{abstract}
The integration of large language models (LLMs) into education offers significant potential to enhance accessibility and engagement, yet their high computational demands limit usability in low-resource settings, exacerbating educational inequities. To address this, we propose an offline Retrieval-Augmented Generation (RAG) pipeline that pairs a small language model (SLM) with a robust retrieval mechanism, enabling factual, contextually relevant responses without internet connectivity. We evaluate the efficacy of this pipeline using domain-specific educational content, focusing on biology coursework. Our analysis highlights key challenges: smaller models, such as SmolLM, struggle to effectively leverage extended contexts provided by the RAG pipeline, particularly when noisy or irrelevant chunks are included. To improve performance, we propose exploring advanced chunking techniques, alternative small or quantized versions of larger models, and moving beyond traditional metrics like MMLU to a holistic evaluation framework assessing free-form response. This work demonstrates the feasibility of deploying AI tutors in constrained environments, laying the groundwork for equitable, offline, and device-based educational tools.
\end{abstract}


\section{Introduction}
In recent years, the integration of large language models (LLMs) into education has shown great potential for enhancing accessibility and engagement in learning~\citep{Kasneci2023LLMsEducation}. However, LLMs that excel in engagement and knowledge metrics~\citep{zhao2024survey} often require significant computational resources. This poses a challenge for low-income and remote communities, where access to reliable internet and high performance computing is limited, exacerbating educational inequities. To address these challenges, we propose the development of an AI tutor utilizing an entirely offline Retrieval Augmented Generation (RAG) pipeline. This approach enables customization and can operate effectively in resource-constrained environments, offering an initial step toward bridging the resource gap and improving equitable access to AI-powered education.

A primary focus in the field has been the development of lightweight LLM architectures designed for deployment on low-resource hardware. Models such as SmolLM~\citep{allal2024SmolLM, allal2024SmolLM2} and TinyLlama~\citep{zhang2024TinyLlama} exemplify this trend by reducing parameter count and leveraging optimized training techniques to maintain competitive performance. Moreover, RAG pipelines have emerged as a powerful technique for integrating external knowledge into LLM responses~\citep{chen2023benchmarkinglargelanguagemodels}. In educational settings, AI based tooling interaction and knowledge retrieval capabilities are critical. Therefore, research has grown steadily on evaluation alternatives. For example, comprehensibility has been assessed using human evaluations, while accuracy is often benchmarked against predefined ground truth responses, from MMLU~\citep{hendrycks2021measuringmassivemultitasklanguage}.

\section{Current Efforts}

Recent advancements in LLMs have significantly improved their performance in natural language understanding and generation. However, deploying these models in resource-constrained environments remains a challenge due to their high computational and memory demands. To address these limitations, the research community has explored several strategies aimed at optimizing LLMs for efficiency while maintaining accuracy and usability. This section highlights current efforts relevant to our goal of building an accessible, CPU-efficient LLM tutor.

\subsection{Efficient LLM Architectures}
A primary focus in the field has been the development of lightweight LLM architectures designed for deployment on low-resource hardware. Models such as SmolLM~\citep{allal2024SmolLM, allal2024SmolLM2} and TinyLlama~\cite{zhang2024TinyLlama} exemplify this trend by reducing parameter count and leveraging optimized training techniques to maintain competitive performance. SmolLM, for instance, achieves remarkable computational efficiency while retaining the ability to generate coherent and contextually relevant responses. Similarly, TinyLlama has demonstrated the feasibility of downsizing LLMs without significant degradation in their ability to understand and respond to natural language queries. These models form the foundation of our exploration into resource-efficient tutoring systems.

\subsection{Retrieval-Augmented Generation Pipelines}
Retrieval-Augmented Generation (RAG) pipelines have emerged as a powerful technique for integrating external knowledge into LLM responses. By retrieving relevant information from a database or document corpus, RAG pipelines ensure that the model's outputs are grounded in factual content. This approach is particularly valuable in educational settings, where accuracy and relevance are critical. Recent implementations of RAG pipelines, such as OpenAI’s and Hugging Face’s frameworks, have shown promise in combining retrieval and generation seamlessly. In this work, we incorporate a RAG pipeline tailored for the OpenStax biology textbook, allowing users to interact with specific textbook content. This ensures that the LLM serves as a reliable and context-aware tutor.

\subsection{Evaluation of LLMs for Education}
The application of LLMs in education has gained significant attention, with studies exploring their potential as tutors, teaching assistants, and content generators. Key metrics in evaluating these systems include accuracy, comprehensibility, and user satisfaction. For example, comprehensibility has been assessed using human evaluations, while accuracy is often benchmarked against predefined ground truth responses, from MMLU\citep{hendrycks2021measuringmassivemultitasklanguage}. Additionally, researchers have explored how well LLMs align their outputs with educational material to avoid introducing misinformation. We plan to adopt similar evaluation metrics in our study, focusing on how well the system adheres to textbook content and how effectively it communicates complex concepts to young adults.

\section{Pipeline Overview}
\begin{figure}
    \centering
    \includegraphics[width=0.9\linewidth]{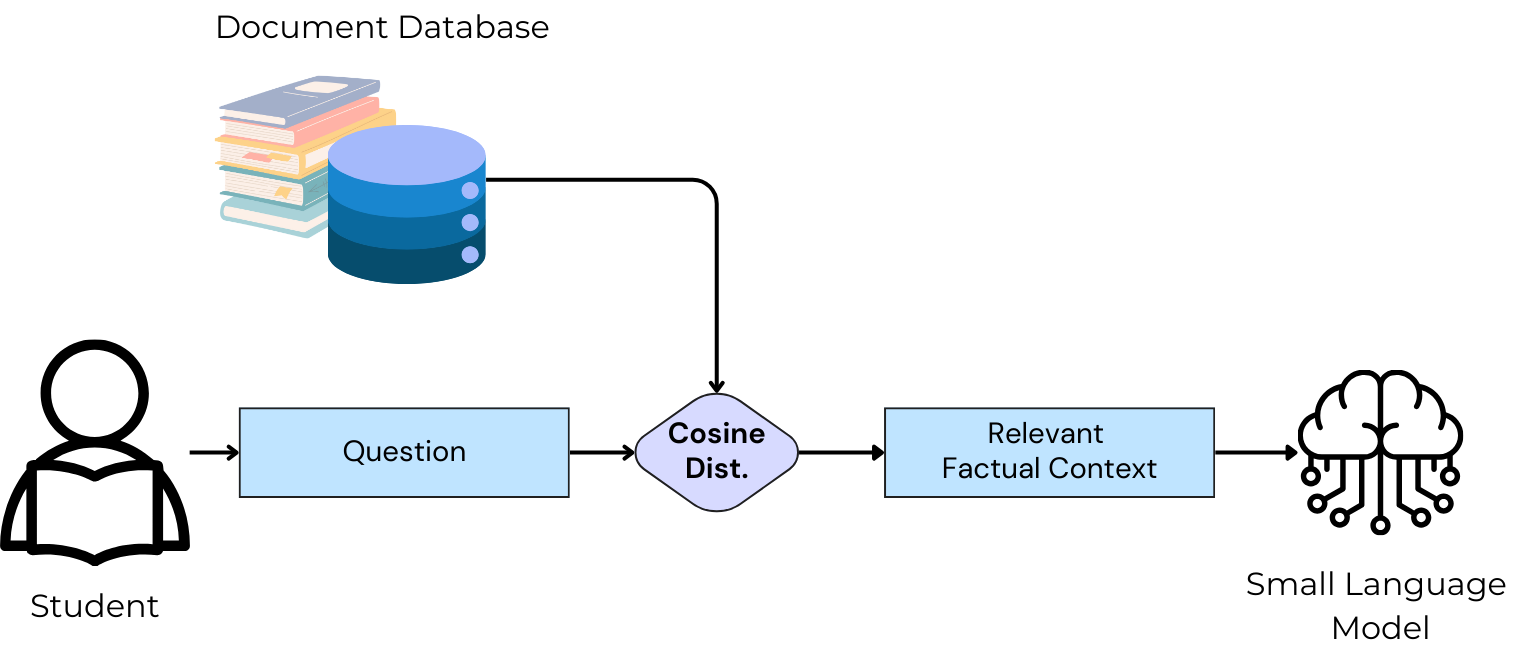}
    \caption{Overview of the AI tutor pipeline.}
    \label{fig:Pipeline}
\end{figure}
The proposed closed system AI Tutor is an offline, computationally efficient system composed of two key components: a RAG pipeline and a small language model (SLM). Figure~\ref{fig:Pipeline} shows an outline of this pipeline. The RAG pipeline ensures factuality by managing knowledge retrieval from a local corpus of the user's choice, using cosine similarity to provide contextually relevant information based on user queries~\citep{wang2020minilmdeepselfattentiondistillation}. Meanwhile, the SLM delivers generative conversational capabilities, enabling engaging and human-like interactions. This design overcomes the limitations of SLMs ~\citep{li2024evaluatingquantizedlargelanguage, jin2024comprehensiveevaluationquantizationstrategies} by augmenting the input prompt with accurate, in-context knowledge, effectively balancing conversational fluency and factual precision.

\subsection{User Input}
An AI Tutor works under conversational setups, where a user can interact through consecutive prompts to the model, dealing with an overarching, single motivation. This single motivation, the conversation's topic, which can vary or feature different degrees of complexity, revolves around a factual question that a student may have regarding the content of a curriculum program for an individual course. Thus, our pipeline is designed on a per-coursework basis (i.e. College Biology) to address any conversation seeking to understand, correct or expand knowledge of a topic of such coursework.

\subsection{Retrieval Augmented Generation}
Given an input prompt from the user, usually in the form of a question seeking a factual answer, our first step is to look for domain specific knowledge that enriches the input to the SLM. This enrichment process consists on the identification of relevant text blocks pieces, retrieved from a pre-loaded vectorized documents database.

This database bears the responsibility of guaranteeing an accurate and grounded response without access to an Internet connection. Thus, it behaves as the pipeline's specialized search engine. Moreover, it can also compensate for the SLM's lower knowledge representation capacity. To build it, we gather a sample of college and high-school level textbooks for the coursework of choice. We naively split these textbooks in text blocks and embed them using FAISS Bert\footnote{\href{https://huggingface.co/sentence-transformers/all-MiniLM-L6-v2}{Embedding model reference}} model.

It is important to note that the database-building step is performed entirely as a pre-processing step. Lowering the computational resources used during interaction with the model. Also, this means that we could eventually use a wide range of preprocessing steps, leveraging higher capacity models like GPT4o-mini of Llama3.3, and less constrained computation specifications, to allow for a better retrieval operation. However, we consider exploration on this track outside the scope of our current analysis.

Once we have a vectorized document database, our pipeline can receive a user prompt, encode it using the same embedded database, and query its largest $K$ closest text fragments from the knowledge base. For our distance computation, we use a traditional cosine distance metric, although there is ample room for exploring alternative approaches to retrieving the most informative chunks based on an input query. In turn, these $K$ text blocks and the initial prompt are formatted in the standard prompt template that properly sets up the task for the language model.

\subsection{Small Language Model}
The formatted prompt is then forwarded to a small language model to generate a human-interpretable factual response. This component is key to ensure we can flexibly adapt to user queries and reply in a helpful, honest, and relatable manner. 

A key characteristic of this pipeline component is its memory footprint. In order to ensure a reliable deployment in computationally constrained environments, we have to work with a small, low-capacity model. This poses a trade-off in terms of the role the language model occupies. If the model is extremely small, it loses context awareness and misses out on generating output that is grounded on the RAG component's context. Moreover, small models can also struggle to deal with complex contexts, either because of their length or due to the indirect relationship between the user's question and the retrieved text blocks content. In turn, a larger model, while more reliable as a conversational agent, will require more computation power and thus higher access barriers for large-scale deployment.

\section{Evaluation}
For our evaluation, we test a range of options for a language model: quantized variations of SmolLM 135M~\citep{allal2024SmolLM} and a larger SmolLM2 1.7B~\citep{allal2024SmolLM2}. To focus on domain-specific evaluation in Biology, we only consider \textit{college} and \textit{high school} Biology tasks in MMLU.

To construct factual support for standard Biology coursework, we construct a database from the textbooks \textit{Biology 2e}~\citep{Clark2018} \textit{Biology AP}~\citep{Rye2016}, \textit{Concepts of Biology}\citep{Fowler2013} from the OpenStax repository \footnote{\href{https://openstax.org/k12/biology}{\texttt{openstax.org/k12/biology}}}. Following the outline in Section 3.2, We use Chroma\footnote{\href{https://github.com/chroma-core/chroma}{\texttt{github.com/chroma-core/chroma}}} as our database of choice and split the knowledge corpora into blocks of size 300 tokens. Moreover, we return the two most similar blocks and append them as context to the question ($k = 2$).

\subsection{Impact of Augmented Context}

\begin{table}[H]
    \centering
    \begin{tabular}{|l|c|c|}
        \hline
        \textbf{Model} & \textbf{\% With RAG} & \textbf{\% W/out RAG} \\
        \hline
        SmolLM (135M) & 20.48 & 20.04 \\ 
        SmolLM2 (135M) & 21.81 & 21.59 \\ 
        SmolLM2 (1.7B) & 33.04 & 41.00 \\ 
        \hline
    \end{tabular}
    \caption{Accuracy Comparison with and without RAG}
    \label{tab:rag_comparison}
\end{table}

Table \ref{tab:rag_comparison} reveals that the inclusion of the RAG component prior to processing user queries does not lead to an improvement in accuracy. On the contrary, it appears to reduce accuracy, with particularly notable declines observed in the larger model. A plausible explanation for this phenomenon is that smaller models may face difficulties in effectively managing extended context windows, such as those created by the formatted prompts that include both the user query and the retrieved textbook blocks.

\subsection{Effect of Context as Noise}
Our initial observed results seem to point at the added context not informing the model. Even worse, such added text appears to be making it harder for the model to parse the whole input and identify an answer. To evaluate this effect from the context, we include the true answers for the multiple choice questions from MMLU in the input to the language model.

We implement different setups:
\begin{itemize}
    \item \textbf{Pure MC answer}: model is fed the MMLU question and an explicit phrase indicating the correct answer to it.
    \item \textbf{RAG Context + MC Answer}: model is fed the same MC answer along with two text blocks retrieved from the RAG component.
\end{itemize}
We also play around with the kind of true multiple-choice answer we show the model, as we experiment with using the option label (a letter $A, B, C$ or $D$) and then using the text option associated with it.

\begin{table}[H]
    \centering
    \begin{tabular}{|l|c|c|}
        \hline
        \textbf{Model} & \textbf{RAG + Letter answer} & \textbf{Letter answer only} \\
        \hline
        SmolLM (135M) & 26.65\% & 91.63\% \\ 
        SmolLM2 (135M) & 40.31\% & 74.89\% \\ 
        SmolLM2 (1.7B) & 73.13\% & 95.81\% \\ 
        \hline
    \end{tabular}
    \caption{Result comparison when providing the letter answer to the model.}
    \label{tab:rag_letter_comparison}
\end{table}

\begin{table}[H]
    \centering
    \begin{tabular}{|l|c|c|}
        \hline
        \textbf{Model} & \textbf{RAG + Text answer} & \textbf{Text answer only} \\
        \hline
        SmolLM (135M) & 20.70\% & 23.35\% \\ 
        SmolLM2 (135M) & 22.91\% & 27.50\% \\ 
        SmolLM2 (1.7B) & 29.50\% & 50.40\% \\ 
        \hline
    \end{tabular}
    \caption{Result comparison when providing the letter answer to the model.}
    \label{tab:rag_text_comparison}
\end{table}

The evaluation framework for MMLU favors providing the correct letter option (e.g., A, B, C, or D) directly, as this simplifies the task for the model and maximizes accuracy. The model assigns a probability to each letter, making it easier to predict correctly when the letter is explicitly given. However, providing the correct answer as text requires the model to associate the text with its corresponding label, making the task more complex and challenging.

As the tables show, in both cases there is a consistent, and significant loss in accuracy when adding the RAG component's output to the model. Contrary to what one would initial expect, the larger model (SmolLM2 1.7B) is as vulnerable to this loss in performance.

\subsection{Effect of High-Quality Retrieval}
\begin{figure}[H]
    \centering
    \includegraphics[width=0.9\linewidth]{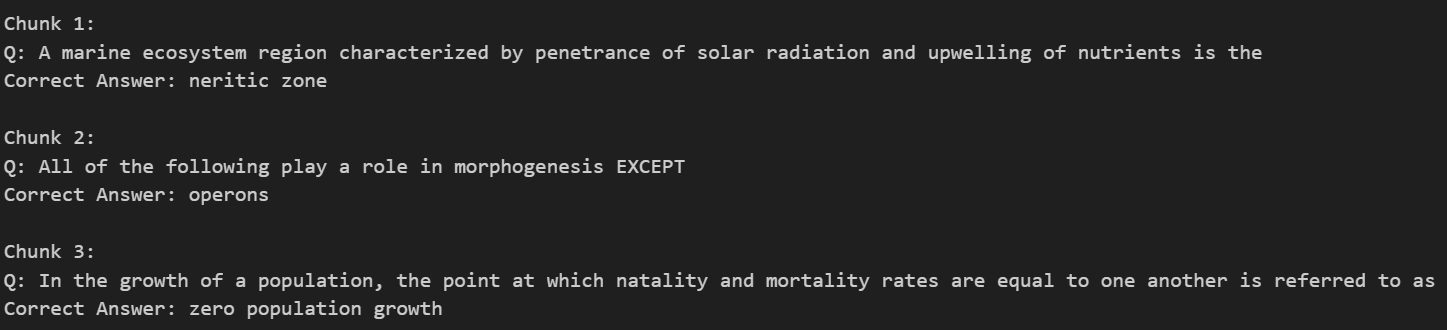}
    \caption{Example of text chunks constructed from MMLU Database}
    \label{fig:MMLU Chunks}
\end{figure}

To evaluate the impact of context content as a potential source of noise, we construct a document database directly from the evaluation dataset (MMLU tasks for high-school and college-level Biology), example chunks shown in Figure~\ref{fig:MMLU Chunks}. We measure model accuracy by feeding questions from the dataset into the pipeline. To ensure a fair comparison, all models are tested using the standard pipeline with context limited to either the single most similar question pair ($k = 1$) or the top two most similar questions ($k = 2$).

\begin{table}[H]
    \centering
    \begin{tabular}{|l|c|c|}
        \hline
        \textbf{Model} & \textbf{MMLU @ k=1} & \textbf{MMLU @ k=2} \\
        \hline
        SmolLM 135M & 24.67\% & 23.35\% \\ 
        SmolLM2 135M & 24.45\% & 24.89\% \\ 
        SmolLM2 1.7B & 58.81\% & 30.84\% \\ 
        \hline
    \end{tabular}
    \caption{Comparison with top 1 and top 2 results when using MMLU as knowledge base}
    \label{tab:top_k_comparison}
\end{table}

The results align with previous observations regarding the detrimental impact of larger contexts on model performance. Notably, even when the retrieved context is highly relevant and closely aligned with the input question, the loss in performance persists. These findings suggest that small models are particularly sensitive to extended contexts, irrespective of their informational quality when handling factual tasks.

\section{Next Steps}
So far, we have focused on identifying the weaknesses in our pipeline to systematically address them. Based on the results highlighted above, we have identified two critical areas for improvement:

1) \textbf{Enhancing chunking to minimize noise:} We aim to explore chunking techniques that produce highly specialized chunks, ensuring that each chunk in the embedded database contains only relevant information for the given query. Promising approaches include semantic chunking~\citep{Suro2024SemanticTokens, Hacioglu2003SemanticChunking}, agentic chunking, and meta chunking~\citep{zhao2024metachunking}. Semantic chunking clusters sentences with similar embeddings to create cohesive units, improving the language model's ability to process and interpret the information. Agentic chunking leverages language models to dynamically segment text based on task-specific nuances, offering greater context adaptability. Meta chunking takes a broader approach by organizing text into logical units spanning multiple sentences or paragraphs, capturing deeper linguistic and contextual relationships. These techniques should improve the quality of information fed into the model.

2) \textbf{Exploring and Customizing Language Models:} While SmolLM is an effective model on its own, our results suggest it struggles to fully leverage the additional information provided by the RAG pipeline. Specifically, the performance drop observed between Table~\ref{tab:rag_letter_comparison} and Table~\ref{tab:rag_text_comparison} implies that SmolLM is not adept at utilizing retrieved information to enhance its responses. This limitation becomes even more evident when there is noise in the chunks, as shown in Table~\ref{tab:top_k_comparison}.

It is essential to determine whether this drawback is unique to SmolLM or indicative of a broader limitation in language models of similar size. To address this, we plan to conduct evaluations with a range of alternative models, including quantized versions of larger models. This will allow us to assess whether scaled-down versions of larger language models are better equipped to handle RAG’s retrieval capabilities compared to originally small models like SmolLM.

As part of this exploration, we must also revisit our evaluation methods. The current metric, MMLU, falls short in capturing the nuanced performance required for fine-tuned or customized models in RAG pipelines. We plan to move to a more holistic evaluation framework that enables the model to generate free-form responses. These responses will then be evaluated for factual accuracy, coherence, and relevance, against larger models. Providing a more comprehensive understanding of the model’s ability to handle real-world queries.

The burden of domain-specific knowledge is on the RAG component of the pipeline and not the model, switching to this evaluation framework would allow us to evaluate the language model on what it is actually responsible for, taking in the user's prompt and related RAG chunks, identifying the most relevant information and formulating an easily comprehensible response. With this in mind, in the future we may explore finetuning current models with this task in mind, or possibly training our own SLM from scratch.

\section{Implications}
With the goal of developing a compact an accessible AI tutor, we believe it's important to keep some end products in mind. Once working, we will adapt the pipeline for a smartphone application, leveraging the rapid advancements in mobile technology over the past 15 years. Modern smartphones are now capable of supporting highly distilled LLMs~\citep{Fassold2024PhoneLLMs}, making them a viable platform for deploying a robust, on-device RAG pipeline. This approach has the potential to bring accessible, AI-driven educational tools directly into users’ hands, enabling a seamless, internet-independent experience.

However, despite the widespread proliferation of smartphones, many resource-limited communities still lack access to modern devices or reliable internet infrastructure. To address this challenge, we propose deploying the RAG pipeline on cost-effective, low-power systems run by a Raspberry Pi 5. These devices can be set up as community knowledge hubs, providing local and offline access to a comprehensive AI tutor. By combining smartphone-based applications for widespread reach with low-cost Raspberry Pi systems for under-served regions, we aim to bridge the digital divide and ensure equitable access to information and education.

\section{Code Availability}
All relevant code is available at \href{https://github.com/JuanseHevia/quantized-education-v2/}{https://github.com/JuanseHevia/quantized-education-v2}. 

\newpage
\bibliographystyle{plain}
\bibliography{bibliography}

\begin{thebibliography}{17}
\providecommand{\natexlab}[1]{#1}
\providecommand{\url}[1]{\texttt{#1}}
\expandafter\ifx\csname urlstyle\endcsname\relax
  \providecommand{\doi}[1]{doi: #1}\else
  \providecommand{\doi}{doi: \begingroup \urlstyle{rm}\Url}\fi

\bibitem[Allal et~al.(2024{\natexlab{a}})Allal, Lozhkov, Bakouch, Blázquez, Tunstall, Piqueres, Marafioti, Zakka, von Werra, and Wolf]{allal2024SmolLM2}
Loubna~Ben Allal, Anton Lozhkov, Elie Bakouch, Gabriel~Martín Blázquez, Lewis Tunstall, Agustín Piqueres, Andres Marafioti, Cyril Zakka, Leandro von Werra, and Thomas Wolf.
\newblock Smollm2 - with great data, comes great performance, 2024{\natexlab{a}}.

\bibitem[Allal et~al.(2024{\natexlab{b}})Allal, Lozhkov, Bakouch, von Werra, and Wolf]{allal2024SmolLM}
Loubna~Ben Allal, Anton Lozhkov, Elie Bakouch, Leandro von Werra, and Thomas Wolf.
\newblock Smollm - blazingly fast and remarkably powerful, 2024{\natexlab{b}}.

\bibitem[Chen et~al.(2023)Chen, Lin, Han, and Sun]{chen2023benchmarkinglargelanguagemodels}
Jiawei Chen, Hongyu Lin, Xianpei Han, and Le~Sun.
\newblock Benchmarking large language models in retrieval-augmented generation, 2023.
\newblock URL \url{https://arxiv.org/abs/2309.01431}.

\bibitem[Clark et~al.(2018)Clark, Douglas, and Choi]{Clark2018}
Mary~Ann Clark, Matthew Douglas, and Jung Choi.
\newblock \emph{Biology 2e}.
\newblock OpenStax, Houston, Texas, 2018.
\newblock URL \url{https://openstax.org/books/biology-2e/pages/1-introduction}.
\newblock Accessed from: \url{https://openstax.org/books/biology-2e/pages/1-introduction}.

\bibitem[Fassold(2024)]{Fassold2024PhoneLLMs}
Hannes Fassold.
\newblock Porting large language models to mobile devices for question answering, 2024.
\newblock URL \url{https://arxiv.org/abs/2404.15851}.

\bibitem[Fowler et~al.(2013)Fowler, Roush, and Wise]{Fowler2013}
Samantha Fowler, Rebecca Roush, and James Wise.
\newblock \emph{Concepts of Biology}.
\newblock OpenStax, Houston, Texas, 2013.
\newblock URL \url{https://openstax.org/books/concepts-biology/pages/1-introduction}.
\newblock Accessed from: \url{https://openstax.org/books/concepts-biology/pages/1-introduction}.

\bibitem[Hacioglu and Ward(2003)]{Hacioglu2003SemanticChunking}
Kadri Hacioglu and Wayne Ward.
\newblock Target word detection and semantic role chunking using support vector machines.
\newblock In \emph{Proceedings of the 2003 Conference of the North American Chapter of the Association for Computational Linguistics on Human Language Technology: Companion Volume of the Proceedings of HLT-NAACL 2003--Short Papers - Volume 2}, NAACL-Short '03, page 25–27, USA, 2003. Association for Computational Linguistics.
\newblock \doi{10.3115/1073483.1073492}.
\newblock URL \url{https://doi.org/10.3115/1073483.1073492}.

\bibitem[Hendrycks et~al.(2021)Hendrycks, Burns, Basart, Zou, Mazeika, Song, and Steinhardt]{hendrycks2021measuringmassivemultitasklanguage}
Dan Hendrycks, Collin Burns, Steven Basart, Andy Zou, Mantas Mazeika, Dawn Song, and Jacob Steinhardt.
\newblock Measuring massive multitask language understanding, 2021.
\newblock URL \url{https://arxiv.org/abs/2009.03300}.

\bibitem[Jin et~al.(2024)Jin, Du, Huang, Liu, Luan, Wang, and Xiong]{jin2024comprehensiveevaluationquantizationstrategies}
Renren Jin, Jiangcun Du, Wuwei Huang, Wei Liu, Jian Luan, Bin Wang, and Deyi Xiong.
\newblock A comprehensive evaluation of quantization strategies for large language models, 2024.
\newblock URL \url{https://arxiv.org/abs/2402.16775}.

\bibitem[Kasneci et~al.(2023)Kasneci, Sessler, Küchemann, Bannert, Dementieva, Fischer, Gasser, Groh, Günnemann, Hüllermeier, Krusche, Kutyniok, Michaeli, Nerdel, Pfeffer, Poquet, Sailer, Schmidt, Seidel, Stadler, Weller, Kuhn, and Kasneci]{Kasneci2023LLMsEducation}
Enkelejda Kasneci, Kathrin Sessler, Stefan Küchemann, Maria Bannert, Daryna Dementieva, Frank Fischer, Urs Gasser, Georg Groh, Stephan Günnemann, Eyke Hüllermeier, Stephan Krusche, Gitta Kutyniok, Tilman Michaeli, Claudia Nerdel, Jürgen Pfeffer, Oleksandra Poquet, Michael Sailer, Albrecht Schmidt, Tina Seidel, Matthias Stadler, Jochen Weller, Jochen Kuhn, and Gjergji Kasneci.
\newblock Chatgpt for good? on opportunities and challenges of large language models for education.
\newblock \emph{Learning and Individual Differences}, 103:\penalty0 102274, 2023.
\newblock ISSN 1041-6080.
\newblock \doi{https://doi.org/10.1016/j.lindif.2023.102274}.
\newblock URL \url{https://www.sciencedirect.com/science/article/pii/S1041608023000195}.

\bibitem[Li et~al.(2024)Li, Ning, Wang, Liu, Shi, Yan, Dai, Yang, and Wang]{li2024evaluatingquantizedlargelanguage}
Shiyao Li, Xuefei Ning, Luning Wang, Tengxuan Liu, Xiangsheng Shi, Shengen Yan, Guohao Dai, Huazhong Yang, and Yu~Wang.
\newblock Evaluating quantized large language models, 2024.
\newblock URL \url{https://arxiv.org/abs/2402.18158}.

\bibitem[Rye et~al.(2016)Rye, Wise, Jurukovski, DeSaix, Choi, and Avissar]{Rye2016}
Connie Rye, Robert Wise, Vladimir Jurukovski, Jean DeSaix, Jung Choi, and Yael Avissar.
\newblock \emph{Biology}.
\newblock OpenStax, Houston, Texas, 2016.
\newblock URL \url{https://openstax.org/books/biology/pages/1-introduction}.
\newblock Accessed from: \url{https://openstax.org/books/biology/pages/1-introduction}.

\bibitem[Suro(2024)]{Suro2024SemanticTokens}
Joel Suro.
\newblock Semantic tokens in retrieval augmented generation, 2024.
\newblock URL \url{https://arxiv.org/abs/2412.02563}.

\bibitem[Wang et~al.(2020)Wang, Wei, Dong, Bao, Yang, and Zhou]{wang2020minilmdeepselfattentiondistillation}
Wenhui Wang, Furu Wei, Li~Dong, Hangbo Bao, Nan Yang, and Ming Zhou.
\newblock Minilm: Deep self-attention distillation for task-agnostic compression of pre-trained transformers, 2020.
\newblock URL \url{https://arxiv.org/abs/2002.10957}.

\bibitem[Zhang et~al.(2024)Zhang, Zeng, Wang, and Lu]{zhang2024TinyLlama}
Peiyuan Zhang, Guangtao Zeng, Tianduo Wang, and Wei Lu.
\newblock {TinyLlama: An Open-Source Small Language Model}, 2024.
\newblock URL \url{https://arxiv.org/abs/2401.02385}.

\bibitem[Zhao et~al.(2024{\natexlab{a}})Zhao, Ji, Feng, Qi, Niu, Tang, Xiong, and Li]{zhao2024metachunking}
Jihao Zhao, Zhiyuan Ji, Yuchen Feng, Pengnian Qi, Simin Niu, Bo~Tang, Feiyu Xiong, and Zhiyu Li.
\newblock Meta-chunking: Learning efficient text segmentation via logical perception, 2024{\natexlab{a}}.
\newblock URL \url{https://arxiv.org/abs/2410.12788}.

\bibitem[Zhao et~al.(2024{\natexlab{b}})Zhao, Zhou, Li, Tang, Wang, Hou, Min, Zhang, Zhang, Dong, Du, Yang, Chen, Chen, Jiang, Ren, Li, Tang, Liu, Liu, Nie, and Wen]{zhao2024survey}
Wayne~Xin Zhao, Kun Zhou, Junyi Li, Tianyi Tang, Xiaolei Wang, Yupeng Hou, Yingqian Min, Beichen Zhang, Junjie Zhang, Zican Dong, Yifan Du, Chen Yang, Yushuo Chen, Zhipeng Chen, Jinhao Jiang, Ruiyang Ren, Yifan Li, Xinyu Tang, Zikang Liu, Peiyu Liu, Jian-Yun Nie, and Ji-Rong Wen.
\newblock {A Survey of Large Language Models}, 2024{\natexlab{b}}.
\newblock URL \url{https://arxiv.org/abs/2303.18223}.

\end{thebibliography}
\end{document}